\documentclass[a4paper]{jpconf}

\usepackage{amsmath,amssymb,amsthm,amsfonts}
\usepackage{graphicx} 
\usepackage{enumerate}
\usepackage{multicol}

\def \nbb{\mathbb{N}}
\def \rbb{\mathbb{R}}

\def \bcal {\mathcal{B}}

\def \ncal {\mathcal{N}}
\def \ocal {\mathcal{O}}

\def \wcal {\mathcal{W}}

\def \bfk  {\mathfrak{B}}

\newcommand{\thetav}{\boldsymbol{\theta}}
\newcommand{\etav}{\boldsymbol{\eta}}

\newcommand{\zetav}{\boldsymbol{\zeta}}
\newcommand{\piv}{\boldsymbol{\pi}}

\newcommand{\zerov}{\boldsymbol{0}}

\def \. { \,\! }

\def\clap#1{\hbox to 0pt{\hss#1\hss}}

\DeclareMathOperator{\im}{Im}

\newcommand{\idop}{\boldsymbol{1}}

\DeclareMathOperator{\supp}{supp}

\def \expltext#1 {\\ \text{\footnotesize{ (#1) }}\\}
\def \intercomm#1 {\\ \text{\footnotesize{ (#1) }}\\}
\def \undercomm#1 {\underset{\text{\scriptsize{ (#1) }}}}
\def \overcomm#1 {\overset{\text{\scriptsize{ (#1) }}}}

\newcommand{\Hil}{\mathcal{H}}

\newcommand{\boundedops}{\bfk(\Hil)}

\newcommand{\A}{\mathcal{A}}

\newcommand{\hscalar}[2]{\langle #1 , #2 \rangle }

\newcommand{\gnorm}[2]{\lVert #1 \rVert_{#2}}

\newcommand{\zd}{z^\dagger}

\DeclareMathOperator*{\res}{res}

\newcommand{\cmponly}[1]{}
\newcommand{\ahponly}[1]{}

\begin{document}
\title{On the status of pointlike fields in integrable QFTs}

\author{Henning Bostelmann}

\address{University of York, Department of Mathematics, York YO10 5DD, United Kingdom}

\ead{henning.bostelmann@york.ac.uk}

\begin{abstract}
In integrable models of quantum field theory, local fields are normally constructed by means of the bootstrap-formfactor program. However, the convergence of their $n$-point functions is unclear in this setting. An alternative approach uses fully convergent expressions for fields with weaker localization properties in spacelike wedges, and deduces existence of observables in bounded regions from there, but yields little information about their explicit form. We propose a new, hybrid construction: We aim to describe pointlike local quantum fields; but rather than exhibiting their $n$-point functions and verifying the Wightman axioms, we establish them as closed operators affiliated with a net of local von Neumann algebras that is known from the wedge-local approach. This is shown to work at least in the Ising model.
\end{abstract} 

\section{Introduction}

Quantum field theory (QFT), describing the behaviour of subatomic particles at relativistic speeds, is usually formulated in terms of the eponymous quantum fields: A quantum field $\Phi(x)$---and we will restrict to scalar Bose fields in all what follows---is a quantum observable localized at the spacetime point $x$, where localization manifests itself in commutativity at spacelike distances:
\begin{equation}\label{eq:localcommute}
    [\Phi(x), \Phi(y)] = 0 \quad \text{if $x$ is spacelike separated from $y$}.
\end{equation}

However, localization at a single point $x$ is an unphysical over-idealization. By Heisenberg's uncertainty relation, one expects a quantity localized arbitrarily sharp in time and space to be arbitrarily \emph{delocalized} in energy and momentum; in other words, these fields should have an infinite energy-momentum transfer. 

From a mathematical perspective, this over-idealization is reflected in singularities of the fields. While often referred to as ``operators on a Hilbert space'' in the physics literature, the $\Phi(x)$ are actually only defined as quadratic forms: Even in the simplest examples, $\Phi(x)\psi$ is never a normalizable vector in the Hilbert space of the theory, regardless of the vector $\psi$, not even if $\psi=\Omega$ is the vacuum vector. Only matrix elements $\hscalar{\psi_1}{ \Phi(x) \psi_2}$ make sense as finite numbers if both $\psi_{1,2}$ are sufficiently ``well-behaved'' vectors, for example, with an energy cutoff. Consequently, products $\Phi(x)\Phi(y)$ of fields do not exist in general (or require some regularization procedure), and there is no notion of spectral projections or (quasi-)eigenvalues of $\Phi(x)$, all of which would be fundamental to a physical interpretation.

The usual remedy to this problem is to average the fields (or, mathematically, to treat them as operator-valued distributions): If $g$ is a smooth, compactly supported function on spacetime, then $\Phi(g):=\int dx \,g(x)\Phi(x)$ can often be defined as an operator, at least on a dense set of vectors. In other words, smearing in spacetime improves the high-energy singularities of the fields, and allows us to define their products etc. This approach, first formalized by Wightman \cite{StrWig:PCT}, is known to work at least in free theories as well as in the interacting $P(\Phi)_2$ and $\Phi_3^4$ models \cite{GliJaf:quantum_physics}.
However, can this procedure work in all QFTs with self-interaction? Even if that were the case in principle, the operator domains of the $\Phi(g)$ are often hard to access in concrete models, and their functional analytic properties remain difficult to obtain.

Here we investigate this problem in a specific, simplified model of interaction, namely integrable QFTs in $1+1$ spacetime dimensions. (We will recall their construction in Sec.~2.) Traditionally, in this situation, one tries to define the fields $\Phi(g)$ or $\Phi(x)$ by specifying their $n$-point functions, which are given as an infinite series. However, convergence of this series is hard to control---yet another reflection of the singular nature of pointlike fields. 

We present an alternative approach: We define the quantum field theory indirectly via a net of local von Neumann algebras associated with spacetime regions. Then, we show that the $\Phi(g)$ exist as (unbounded) operators that are \emph{affiliated} with the algebras (Sec.~\ref{sec:joint}), thus bypassing the treatment of products of fields or of their $n$-point functions. In this way, we ameliorate (though not eliminate) the convergence problem of the series, breaking it down to a problem that is tractable at least in the simplest example, the massive Ising model.

The present paper contains a brief and mostly non-technical summary of these results; the reader is referred to \cite{BostelmannCadamuro:examples} for details.

\section{Construction of integrable models}

Integrable QFTs are quantum field theories on 1+1 dimensional Minkowski space with a simplified kind of self-interaction: They describe particles under elastic scattering on the two-particle level only; scattering between an arbitrary number of particles is merely a combination of two-particle scattering processes, i.e., the model has a factorizing scattering matrix. There is no particle creation or annihilation; the particle number as well as the momenta of individual particles are conserved at all times. Hence, in order to specify such a theory, it is enough to know the particle spectrum (masses and spin of elementary particles) and the scattering matrix at the two-particle level. In the present article, we focus on one species of scalar particle of mass $\mu>0$, with no bound states. The only input then required is the two-particle scattering function $S$. In this section, we recall how a corresponding quantum field theory is constructed from this data.

\subsection{General setup}

Our model is specified by the particle mass $\mu$ and a two-particle scattering function $S$, that is, a meromorphic function which is analytic and bounded in the physical strip $\rbb+i[0,\pi]$ and fulfills 
\begin{equation}
S(\zeta)^{-1}=S(-\zeta)=\overline{S(\bar{\zeta})\vphantom{\hat S}}=S(\zeta+i \pi). 
\end{equation}
Examples include the massive Ising model, where $S=-1$, and the sinh-Gordon model, where $S(\zeta)=\frac{\sinh \zeta - i \sin B \pi/2 }{\sinh \zeta + i \sin B \pi/2 }$ with some $B\in(0,1)$.

Given such function $S$, one defines a representation of the Zamolodchikov-Faddeev algebra, a modified CCR algebra with creators and annihilators $\zd(\theta)$, $z(\theta)$, depending on rapidities $\theta$, which fulfill
\begin{align}
 z^\#(\theta_1)    z^\#(\theta_2) &= {S(\theta_1-\theta_2)} \, z^\#(\theta_2) z^\#(\theta_1)\, & (z^\# = \zd,z),\\
 z(\theta_1)   \zd(\theta_2) &= {S(\theta_2-\theta_1)} \,\zd(\theta_2)z(\theta_1)+\delta(\theta_1-\theta_2)\cdot \idop.
\end{align}
These act on a modified Fock space $\Hil$ spanned by $n$-particle vectors of the form
\begin{equation}\label{eq:npv}
   \psi_n = \int d^n\theta \, f(\theta_1,\ldots,\theta_n)\, \zd(\theta_1)\dotsm \zd(\theta_n)\Omega,
\end{equation}
where $\Omega$ is the Fock vacuum. If the support of $f$ is such that the rapidities are in ascending, respectively descending, order, then $\psi_n$ can be interpreted as an outgoing, respectively incoming, particle configuration with wave function $f$, although this interpretation can be justified only after introducing local observables \cite[Sec.~6]{Lechner:2008}.

Spacetime symmetries are represented on $\Hil$ as follows: 
translations $T_x$ (for $x \in \rbb^2$) and boosts $B_\lambda$ (for $\lambda \in \rbb$) are fixed by the relations 
 \begin{align}
  U(T_x) \zd(\theta) U(T_x)^\ast = e^{ip(\thetav)\cdot x} \zd(\theta), 
  \qquad U(B_\lambda) \zd(\theta) U(B_\lambda)^\ast = \zd(\theta+\lambda)  
 \end{align}
and $U(T_x)\Omega=U(B_\lambda)\Omega=\Omega$; here $p(\thetav)=\mu(\cosh\theta,\sinh\theta)$. Most importantly, spacetime reflections $R$ act on  $\psi_n$ (as in \eqref{eq:npv}) by  
\begin{equation}
 U(R)\psi_n = \int d^n\theta \, \overline{f(\theta_1,\ldots,\theta_n)}\, \zd(\theta_n)\dotsm \zd(\theta_1)\Omega.
\end{equation}
The generator of time translations, i.e., the Hamiltonian $H$, is thus given by
\begin{equation}\label{eq:hamil}
 H\psi_n = \int d^n\theta \,\Big(\sum_{j=1}^n \mu \cosh \theta_j \Big) f(\theta_1,\ldots,\theta_n)\, \zd(\theta_1)\dotsm \zd(\theta_n)\Omega.
\end{equation}

\subsection{Form factor program}

In order to construct quantum fields, the longest established approach is the \emph{form factor program} (see \cite{BabujianFoersterKarowski:2006} for a review).
Here one makes an \emph{ansatz} for the local fields as follows. One formally expands their $n$-point functions via an intermediate basis of sharp-rapidity states, for example for $n=2$,
   \begin{equation}\label{eq:ffseries}
    \hscalar{\Omega}{\Phi(x)\Phi(y)\Omega} = \sum_{k=0}^\infty \int \frac{d\theta_1 \dotsm d\theta_k}{k!} \big\lvert 
    \underbrace{\langle \zd(\theta_1)\dotsm \zd(\theta_k)\Omega |\Phi(0)  \Omega \rangle}_{=:F_k(\thetav)} 
    \big\rvert^2 e^{i(y-x)\cdot\sum_j p(\theta_j)} \;.
  \end{equation}
  It turns out that the expansion coefficients $F_k$, called the \emph{form factors} (for which we use slightly different conventions than usual), already determine all $n$-point functions. The expected properties of the field $\Phi$, such as locality and covariance under $U$, lead to restrictions on the $F_k$, the \emph{form factor equations}. Namely, the $F_k$ have meromorphic continuations, analytic where $\im \zeta_1 < \ldots < \im \zeta_k < \im \zeta_1 + \pi$ with at most first-order poles at the boundary of this region, and they fulfill, $1 \leq j < k$,
  \begin{align} \label{eq:ssymm}
F_k(\zetav)
&=  S(\zeta_{j+1}-\zeta_j) F_k(\zeta_1,\dotsc,\zeta_{j+1},\zeta_j,\dotsc,\zeta_k),
 \\ \label{eq:speriod}
 F_k (\zetav)  &= F_k (\zeta_2,\dotsc,\zeta_{k},\zeta_1+2\pi i), 
 \\  \label{eq:kinres}
\res_{\zeta_2-\zeta_1 = i \pi} F_{k}( \zetav )
&= - \frac{1}{2\pi i }
\Big(1-\prod_{j=1}^{k} S(\zeta_1-\zeta_j) \Big)
F_{k-2}( \zeta_3,\dotsc,\zeta_k ).
  \end{align}
  Given a specific $S$, explicit solutions of these equations can be found in relevant cases. 

  However, now returning to the infinite series \eqref{eq:ffseries}, only very partial results are known about its convergence \cite{BabujianKarowski:2004}, which remains the open mathematical point in the construction.
  
\subsection{Algebraic construction}\label{sec:algebraic}

\begin{figure}
\begin{multicols}{2}
 
\begin{center}
\hphantom{X}

 \includegraphics[height=0.2\textheight]{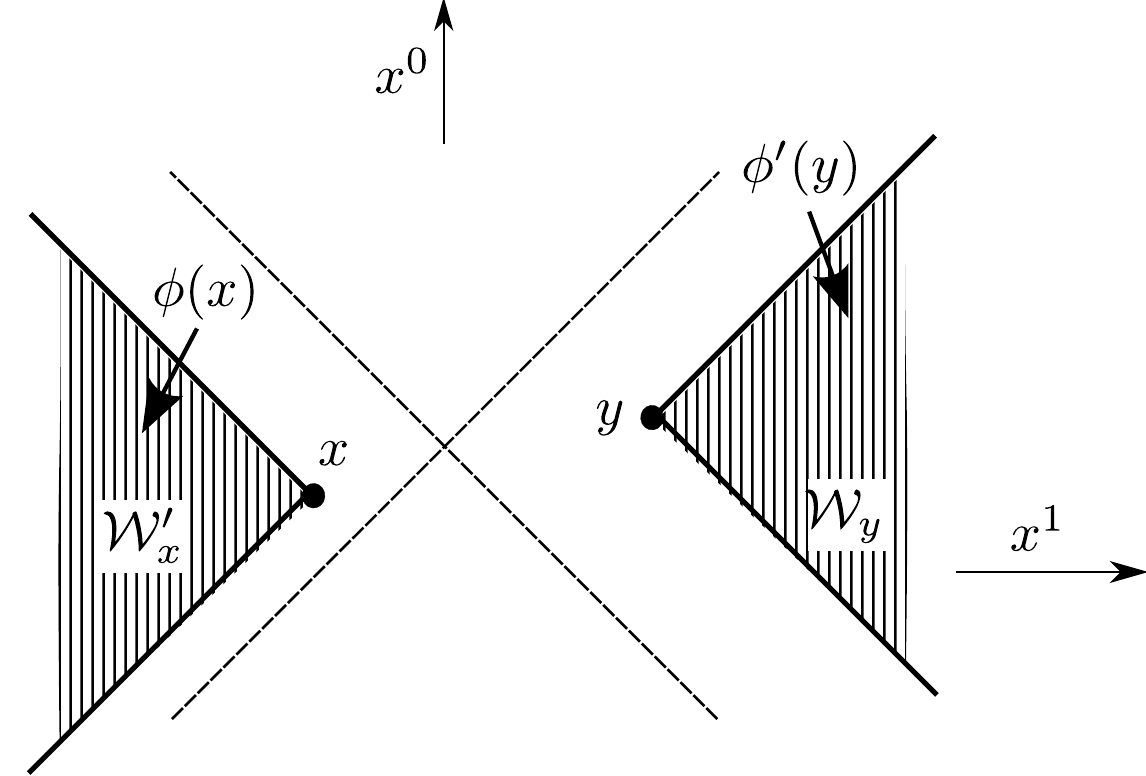}
 \vspace{0.3cm}
 \caption{Fields localized in spacelike wedges}\label{fig:wedges}
\end{center}
\columnbreak
\begin{center}
 \includegraphics[height=0.23\textheight]{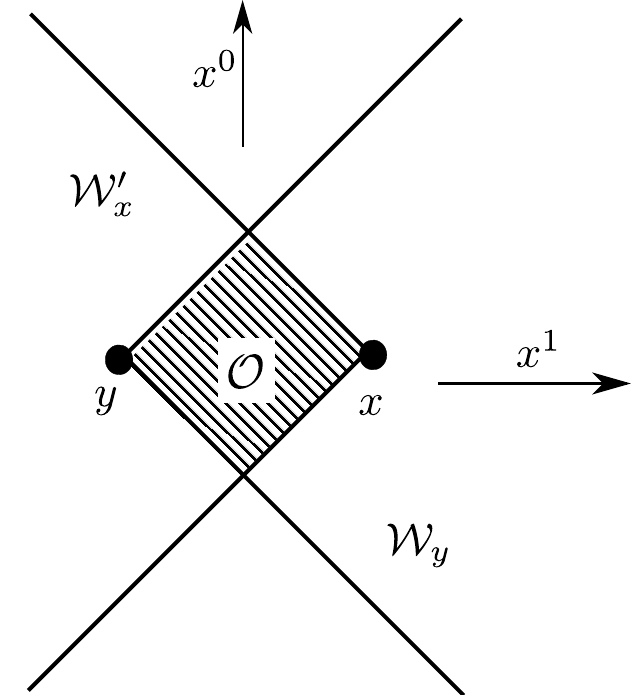}
 \caption{Intersection of two wedges}\label{fig:dc}
\end{center}
\end{multicols}
\end{figure}

An alternative approach, initiated by Schroer \cite{SchroerWiesbrock:2000-1}, uses an indirect route to describe local observables. We temporarily depart from pointlike localized objects; as already seen for smeared fields $\Phi(g)$, weaker localization properties lead to milder high-energy singularities. An even less strictly localized object is the following field:
\begin{equation}
  \phi(x) := \int d\theta \, \Big(e^{ip(\theta) \cdot x} \zd(\theta) + e^{-ip(\theta) \cdot x} z(\theta) \Big).
\end{equation}
This $\phi$ is \emph{not} local in the sense of Eq.~\eqref{eq:localcommute}. However, with $\phi'(x) := U(R) \phi(-x) U(R)$, one finds
\begin{equation}
   [\phi(x),\phi'(y)] = 0  \quad \text{ if  $x$ is spacelike separated \emph{to the left} of } y.
\end{equation}
This can usefully be interpreted as follows: $\phi(x)$ is localized in the spacelike \emph{wedge} $\wcal_x'$ with tip at $x$, opening to the left, and $\phi'(y)$ is localized in the wedge $\wcal_y$ with tip at $y$, opening to the right; the fields commute if the wedges are spacelike separated (see Fig.~\ref{fig:wedges}).

As the main advantage of these wedge-local quantities, $\phi$ is easy to control mathematically---it behaves almost like a free field and can be treated with same methods \cite[Sec.~X.7]{ReedSimon:1975-2}. In particular, the smeared field $\phi(g)$ is essentially selfadjoint, hence one can pass to its bounded functions $\exp i \phi(g)$, $\supp g \subset \wcal_x'$. We call $\A(\wcal_x')$ the von Neumann algebra generated by all $\exp i \phi(g)$ with $\supp g \subset \wcal_x'$, and analogously we define $\A(\wcal_y)$. Importantly, they fulfill
\begin{equation}\label{eq:wedgedual}
    \A(\wcal_x') = \A(\wcal_x)',
\end{equation}
the prime on the right-hand side denoting the commutant of the von Neumann algebra.

Now for a bounded region $\ocal = \wcal_x' \cap \wcal_y$ as in Fig.~\ref{fig:dc}, set 
 \begin{equation}\label{eq:algsect}
\A(\ocal):=\A(\wcal_x')\cap \A(\wcal_y).  
 \end{equation}
 This algebra $\A(\ocal)$ then contains strictly (though not pointlike) local operators. One can check without much effort that these algebras are indeed local and covariant in the sense of the Haag-Kastler axioms \cite{Haa:LQP}. It is not so immediate that $\A(\ocal)$ contains any operator except multiples of the identity; but at least for certain $S$ and for sufficiently large regions, one can establish that $\A(\ocal)\Omega$ is indeed dense in $\Hil$ \cite{Lechner:2008,AL2017}.

This approach solves all mathematical convergence problems, but it has a quite different shortcoming: the explicit form of the local operators $A \in \A(\ocal)$ remains unclear. Their matrix elements $\hscalar{ \zd(\theta_1)\dotsm \zd(\theta_k)\Omega }{A  \Omega}$ do fulfill the form factor equations \cite{BostelmannCadamuro:characterization}; but in the end, these operators are ``constructed'' using the axiom of choice, and no further information about their relation to pointlike fields or other generators is available.

\section{A combined approach}\label{sec:joint}

From the two approaches to local observables described in the previous section, it seems that we face a binary choice: We can either describe pointlike fields explicitly, but lose control of convergence issues at high particle numbers; or we can fully treat all convergence issues to obtain local operators, but have no access to their explicit form.

The aim of this section is to present a construction that combines aspect of those two approaches, and leads to fully convergent but explicitly accessible observables. 
We are going to define these using the explicit expressions for pointlike fields known from the form factor program. However, we want to interpret them differently: Our aim is not to compute the $n$-point functions of the fields $\Phi(x)$, nor to investigate whether products of two fields $\Phi(g_1)\Phi(g_2)$ exist. Rather, we make use of the underlying (abstract) von Neumann algebras, and show that the $\Phi(g)$ are \emph{affiliated} with the local algebra $\A(\ocal)$ if $\supp g \subset \ocal$. Here being affiliated (the precise definition will be recalled in Sec.~\ref{sec:affiliated} below) is the closest possible way to say that $\Phi(g)$ is an element of $\A(\ocal)$: It cannot literally be contained in the algebra, since $\Phi(g)$ is unbounded but $\A(\ocal)$ consists of bounded operators; but when writing $A = V \int \lambda \,dP(\lambda)$ in polar decomposition, one demands that $V$ and all $P(\lambda)$ are contained in $\A(\ocal)$.  

For obtaining this result, we will proceed as follows.
 \begin{enumerate}[(1)]
  \item Define the field $\Phi(g)$ as a quadratic form. To this end, we use the well-known solutions of the form factor equations;
   \item Show that $\Phi(g)$ is local. Here locality will be defined relative to the wedge-local field $\phi$;
   \item Show that $\Phi(g)$ is closable. That is, we want to extend both $\Phi(g)$ and its adjoint to operators on a dense domain, removing one of the particle number cutoffs. It is at this point that convergence issues of an infinite series show up in the construction;
   \item Show that $\Phi(g)$ is affiliated with $\A(\ocal)$. This will follow from locality (2) and closability (3) by an abstract argument.
  \end{enumerate}

  \noindent
  In the remainder of this section, we will comment briefly on each of these four aspects. Details of the construction can be found in \cite{BostelmannCadamuro:examples}.

\subsection{Definition of fields}

Our starting point are the meromorphic functions $F_k$ which are solutions of the form factor equations \eqref{eq:ssymm}--\eqref{eq:kinres}. Let us give an example for the massive Ising model (i.e., for the case $S=-1$): There, for the basic field or order parameter $\Phi(g)$, the form factors are given by
\begin{equation}\label{eq:ffising}
F_{2k+1}(\zetav) := \frac{1}{(2\pi i)^k} \tilde g(p(\zetav))  
\displaystyle\prod_{1 \leq i < j \leq 2k+1} \!\!\!\!\tanh\frac{\zeta_i-\zeta_j}{2}\; ,
\qquad F_{2k}(\zetav) = 0 . 
\end{equation}
Compared with \cite{SchroerTruong:1978,BergKarowskiWeisz:1979}, note the extra factor $\tilde g(p(\zetav))$ resulting from averaging in configuration space. Also, it is evident from \eqref{eq:ffising} that the $F_{2k+1}$ have poles at $\zeta_j - \zeta_i = i\pi$, the so-called \emph{kinematic poles} required by Eq.~\eqref{eq:kinres}; they will play a role in the following.

Given such a set of form factors $F$, we can then define our field $\Phi(g)$ as (cf.~\cite{Lashkevich:1994,BostelmannCadamuro:expansion})
 \begin{equation}\label{eq:phisum}
 \Phi(g) := \!\!\sum_{m,n=0}^{\infty} \int \frac{d^m\theta d^n\eta}{m!n!} F_{m+n}(\boldsymbol{\theta}+i\boldsymbol{0},\boldsymbol{\eta}+i\boldsymbol{\pi}-i\boldsymbol{0}) 
z^{\dagger}(\theta_1) \dotsm z^{\dagger}(\theta_m) z(\eta_1) \dotsm z(\eta_n) .
\end{equation}
This gives us $\Phi(g)$ as a quadratic form, with well-defined matrix elements $\hscalar{\psi_1}{\Phi(g)\psi_2}$ if $\psi_j$ are ``suitable'' vectors: First, they should both include a particle number cutoff, so that the sum \eqref{eq:phisum} is actually finite in matrix elements; hence no convergence issues arise at this point. Second we demand that they decay sufficiently fast in momentum space; technically, we require that $\lVert \exp(H^{\alpha})\psi_j \rVert < \infty$ where $H$ is the Hamiltonian \eqref{eq:hamil}, and $\alpha\in(0,1)$ is fixed.
 
\subsection{Locality}

We need to establish that our fields are local in a suitable sense. This cannot be done in the usual way \eqref{eq:localcommute}, since products of fields $\Phi(g)$, and hence their commutators, are not defined at this point. Instead, we consider locality relative to the wedge-local field $\phi$; we want to establish that  
\begin{equation} \label{eq:relativelocal}
   [\phi(h_L), \Phi(g)] = 0 = [\phi'(h_R), \Phi(g)]
 \end{equation}
 when $h_L$ is supported spacelike to the left and $h_R$ to the right of $g$. Since $\phi$ changes the particle number only by one, and hence preserves particle-number cutoffs, the commutators in \eqref{eq:relativelocal} are well-defined in matrix elements for suitable $h_L$, $h_R$. That the commutators do actually vanish is then quite direct to verify; it is a consequence of the form factor equations, see \cite[Secs.~4.3 and 5.3]{BostelmannCadamuro:characterization}.

\subsection{Closability}

We now want to pass from a quadratic form $\Phi(g)$, where only matrix elements $\hscalar{\psi_1}{\Phi(g) \psi_2}$ are defined, to an (unbounded) operator on a dense domain, that is, $\Phi(g) \psi$ should be a normalizable vector for vectors $\psi$ from a dense set. In other words, we wish to remove the particle cutoff and energy damping from one side of the matrix element. 
In mathematical terms, we are aiming at a densely defined, closed operator (cf.~\cite[Ch.~3.5]{Kat:perturbation}) that extends $\Phi(g)$, with a corresponding extension for its adjoint; see \cite[Def.~3.1]{BostelmannCadamuro:examples} for the exact technical conditions.

In this step, convergence aspects of the infinite sum \eqref{eq:phisum} play a role. A sufficient condition on closability (i.e., the existence of such extension) turns out to be that 
\begin{equation}\label{eq:summation}
    \sum_{m=0}^\infty \frac{2^{m/2}}{\sqrt{m!}} \Big( \lVert F_{m+n} \rVert_{m \times n}^{(\alpha)} + \lVert F_{m+n} \rVert_{n \times m}^{(\alpha)} \Big) < \infty
\end{equation}
for all $n \in \nbb$, where
\begin{equation}
\gnorm{F_{m+n}}{m \times n}^{(\alpha)} := \frac{1}{2} \lVert T \exp(-H^\alpha) \rVert + \frac{1}{2} \lVert  \exp(-H^\alpha) T \rVert  
\end{equation}
and $T$ is the integral operator with kernel $F_{m+n}(\thetav+i\zerov,\etav+i\piv-i\zerov)$, $\thetav\in\rbb^m$, $\etav\in\rbb^n$.

This summation condition can in fact be verified in the example \eqref{eq:ffising} from the Ising model; to that end, the test function $g$ needs to be chosen such that $\tilde g$ and its derivatives decay faster than $\exp(-\|p\|^{\alpha})$ in momentum space, i.e., $g$ needs to be of Jaffe class \cite{Jaffe:1967}. 
A main difficulty in establishing \eqref{eq:summation} are the kinematical poles of the form factors: They mean that the kernels $F_{m+n}(\thetav+i\zerov,\etav+i\piv-i\zerov)$ are boundary values of meromorphic functions at a pole, i.e., the $T$ are singular integral operators of non-convolution type. Estimating the norms $\lVert F_{m+n} \rVert_{m \times n}^{(\alpha)}$ therefore requires considerable technical effort; see  \cite[Sec.~5.2]{BostelmannCadamuro:examples} for the techniques used.

\subsection{Affiliation}\label{sec:affiliated}

Our last step is to establish a relation between the operator $\Phi(g)$ and the abstractly defined von Neumann algebras $\A(\ocal)$, as introduced in Sec.~\ref{sec:algebraic}. As mentioned, this is by way of affiliation.

We recall the definition: Let $A$ be a closed operator on a dense domain $D(A)\subset\Hil$, and let $\ncal \subset \boundedops$ be a von Neumann algebra. Since $A$ is closed, it can be written in its polar decomposition: $A = V |A| = V \int \lambda \,dP(\lambda)$, where $V$ is a partial isometry, and $|A| = \int \lambda \,dP(\lambda)$ is a positive operator written in spectral decomposition, with spectral projectors $P(\lambda)$. Now $A$ is said to be affiliated with $\ncal$ if both $V$ and all $P(\lambda)$ are elements of $\ncal$.

Equivalently \cite[Thm.~3.16]{JorgensenTian:nca}, $A$ is affiliated with $\ncal$ if the following holds: For every operator $B \in \ncal'$ (the commutant of $\ncal$), we have
\begin{equation}\label{eq:commutcrit}
   B D(A) \subset D(A) \qquad \text{and } AB\psi = BA \psi \;\text{ for all }\psi\in D(A).
\end{equation}

Thus, ``$A$ is affiliated with $\ncal$'', for unbounded operators $A$, generalizes the notion ``$A$ is an element of $\ncal$'' in two equivalent ways: First, its polar data are elements of $\ncal$; second, it commutes with the commutant of $\ncal$.

The task is now to deduce that our field operators $\Phi(g)$ are actually affiliated with $\A(\ocal)$, where $\supp g \subset \ocal$. This turns out to be a consequence of locality (relative to the wedge-local field) and closability. The argument is roughly as follows; details can be found in \cite[Sec.~3]{BostelmannCadamuro:examples}.

Let $\ocal = \wcal_x' \cap \wcal_y$ as in Fig.~\ref{fig:dc} and $\supp g \subset \ocal$. We restrict our attention to the Ising model, where the wedge-local fields $\phi(h)$ are in fact bounded operators; they generate the algebras $\A(\wcal_y')$ if $h$ varies over all test functions with $\supp h \subset \wcal_y'$.  It follows from locality \eqref{eq:relativelocal} that $[\Phi(g),\phi(h)] = 0$, at least in matrix elements. Closability of $\Phi(g)$ allows us to maintain this relation in the sense of operators on the domain of the unbounded operator $\Phi(g)$. By criterion \eqref{eq:commutcrit}, $\Phi(g)$ is then affiliated with $\A(\wcal_y')'=\A(\wcal_y)''=\A(\wcal_y)$; we have used \eqref{eq:wedgedual} here. By a similar reasoning, $\Phi(g)$ is affiliated with $\A(\wcal_x')$ as well. It is thus affiliated with $\A(\wcal_y) \cap \A(\wcal_x')$, which equals $\A(\ocal)$ by Eq.~\eqref{eq:algsect}.

Outside the Ising model, where $\phi(h)$ is not bounded, the argument is somewhat more complicated (see \cite[Prop.~3.5]{BostelmannCadamuro:examples}). One needs to work with the bounded generators $\exp i \phi(h)$, which in turn can be approximated by polynomials in the $\phi(h)$. Indeed, an extra technical assumption is needed here, since it is not clear a priori that $B=\exp i \phi(h)$ maps the domain of $A=\Phi(g)$ into itself, as required for criterion \eqref{eq:commutcrit}.

\section{Conclusions and outlook}

As we have summarized in this paper (and as is laid out in more detail in \cite{BostelmannCadamuro:examples}), we have established the existence of smeared fields $\Phi(g)$ in the Ising model as closed operators affiliated with the local algebras $\A(\ocal)$. This was done for the basic field (order parameter), but the same arguments apply to descendant fields, which are obtained by multiplying the form factors \eqref{eq:ffising} with symmetric Laurent polynomials in the variables $e^{\zeta_j}$. 

This construction bypasses the convergence problems of $n$-point functions that are inherent in the form factor program, and replaces them with the summability criterion \eqref{eq:summation} which can in fact be verified rigorously. On the other hand, we still retain explicit control about the matrix elements of our local observables.

The existence of these fields in the Ising model is maybe not very surprising; the same model has been constructed both in a Euclidean setting \cite{PalmerTracy:Ising} and in the algebraic framework \cite{Lechner:2005}. Our point is rather a conceptual one: It is possible, and technically feasible, to treat the model directly in Minkowski space, giving sense to the local fields \emph{without} referring to their $n$-point functions. 

Whether the same construction applies for more general choices of $S$, for example in the sinh-Gordon model \cite{FringMussardoSimonetti:1993}, remains a point for further investigation. Our present results certainly give hope that the problem could be settled also in this case, replacing the convergence problem of the $n$-point functions with a milder one to show affiliation with the local algebras. However, additional technical difficulties remain to be overcome. Not only is the structure of the form factors increasingly intricate, complicating the estimates, but as mentioned in Sec.~\ref{sec:affiliated}, additional criteria on the domain of the field operators would need to be verified.

\section*{References}

\bibliographystyle{iopart-num} 
\bibliography{integrable}

\providecommand{\newblock}{}
\begin{thebibliography}{10}
\expandafter\ifx\csname url\endcsname\relax
  \def\url#1{{\tt #1}}\fi
\expandafter\ifx\csname urlprefix\endcsname\relax\def\urlprefix{URL }\fi
\providecommand{\eprint}[2][]{\url{#2}}

\bibitem{StrWig:PCT}
Streater R~F and Wightman A~S 1964 {\em PCT, Spin and Statistics, and All
  That\/} (New York: Benjamin)

\bibitem{GliJaf:quantum_physics}
Glimm J and Jaffe A 1987 {\em Quantum Physics -- A functional integral point of
  view\/} 2nd ed (New York: Springer)

\bibitem{BostelmannCadamuro:examples}
Bostelmann H and Cadamuro D 2019 {\em Ann. H. Poincar\'e\/} (to appear)
  (\textit{Preprint} \eprint{arXiv:1806.00269})

\bibitem{Lechner:2008}
Lechner G 2008 {\em Commun. Math. Phys.\/} {\bf 277} 821--60 (\textit{Preprint}
  \eprint{math-ph/0601022})

\bibitem{BabujianFoersterKarowski:2006}
Babujian H~M, Foerster A and Karowski M 2006 {\em SIGMA\/} {\bf 2} 082
  (\textit{Preprint} \eprint{hep-th/0609130})

\bibitem{BabujianKarowski:2004}
Babujian H~M and Karowski M 2004 {\em Int. J. Mod. Phys.\/} {\bf A19S2} 34--49
  (\textit{Preprint} \eprint{hep-th/0301088})

\bibitem{SchroerWiesbrock:2000-1}
Schroer B and Wiesbrock H~W 2000 {\em Rev. Math. Phys.\/} {\bf 12} 301--26
  (\textit{Preprint} \eprint{hep-th/9812251})

\bibitem{ReedSimon:1975-2}
Reed M and Simon B 1975 {\em Methods of Modern Mathematical Physics\/} vol II:
  Fourier Analysis, Self-Adjointness (New York: Academic Press)

\bibitem{Haa:LQP}
Haag R 1996 {\em Local Quantum Physics\/} 2nd ed (Berlin: Springer)

\bibitem{AL2017}
Alazzawi S and Lechner G 2017 {\em Commun. Math. Phys.\/} {\bf 354} 913--56
  (\textit{Preprint} \eprint{arXiv:1608.02359})

\bibitem{BostelmannCadamuro:characterization}
Bostelmann H and Cadamuro D 2015 {\em Commun. Math. Phys.\/} {\bf 337}
  1199--240 (\textit{Preprint} \eprint{arXiv:1402.6127})

\bibitem{SchroerTruong:1978}
Schroer B and Truong T~T 1978 {\em Nucl. Phys.\/} {\bf B144} 80--122

\bibitem{BergKarowskiWeisz:1979}
Berg B, Karowski M and Weisz P 1979 {\em Phys. Rev. D\/} {\bf 19} 2477--9

\bibitem{Lashkevich:1994}
Lashkevich M~Y 1994 Sectors of mutually local fields in integrable models of
  quantum field theory arXiv:hep-th/9406118

\bibitem{BostelmannCadamuro:expansion}
Bostelmann H and Cadamuro D 2013 {\em Journal of Physics A\/} {\bf 46} 095401
  (\textit{Preprint} \eprint{arXiv:1208.4763})

\bibitem{Kat:perturbation}
Kato T 1984 {\em Perturbation Theory for Linear Operators\/} 2nd ed (New York:
  Springer)

\bibitem{Jaffe:1967}
Jaffe A~M 1967 {\em Phys. Rev.\/} {\bf 158}(5) 1454--61

\bibitem{JorgensenTian:nca}
Jorgensen P and Tian F 2017 {\em Non-commutative Analysis\/} (Singapore: World
  Scientific)

\bibitem{PalmerTracy:Ising}
Palmer J and Tracy C 1983 {\em Advances in Applied Mathematics\/} {\bf 4}
  46--102

\bibitem{Lechner:2005}
Lechner G 2005 {\em J. Phys. A: Math. Gen.\/} {\bf 38} 3045 (\textit{Preprint}
  \eprint{math-ph/0405062})

\bibitem{FringMussardoSimonetti:1993}
Fring A, Mussardo G and Simonetti P 1993 {\em Nucl. Phys.\/} {\bf B393} 413--41
  (\textit{Preprint} \eprint{hep-th/9211053})

\end{thebibliography}

\end{document}